\documentclass[aps,prl,twocolumn,superscriptaddress,amsmath,amssymb]{revtex4}
\usepackage{graphicx}
\usepackage{dcolumn}
\usepackage{bm}
\usepackage{color}
\usepackage{txfonts}
\usepackage{comment}

\begin{document}

\title{Susceptible-infected-susceptible model on  networks with eigenvector localization }

\author{ Zong-Wen Wei }
\email{wbravo@mail.ustc.edu.cn}
\affiliation{ Guangdong Province Key Laboratory of Popular High Performance Computers, College of Computer Science and Software Engineering, Shenzhen University, Shenzhen 518060,  China}

\author{Bing-Hong Wang }
\email{bhwang@ustc.edu.cn}
\affiliation{ Department of Modern Physics, University of Science and Technology of China, Hefei 230027,  China}

\date{\today}

\begin{abstract}
It is a longstanding debate on the absence of threshold for susceptible-infected-susceptible (SIS) model on networks with finite second order moment of degree distribution.
The eigenvector localization of the adjacency matrix for a network gives rise to the inactive Griffiths phase featuring slow decay of the activity localized around highly connected nodes due to the dynamical fluctuation.
We show how it dramatically changes our understanding of SIS model,
opening up new possibilities for the debate.
We derive the critical condition for Griffiths to active phase transition: on average, an infected node can further infect another one in the characteristic lifespan of the star subgraph composed of the node and its nearest neighbors.
The system approaches the critical point of avoiding the irreversible dynamical fluctuation and the trap of absorbing state. As a signature of the phase transition, the infection density of a node is not only proportional to its degree, but also proportional to the exponentially growing lifespan of the star.
And the divergence of the average lifespan of the stars is responsible for the vanishing threshold in the thermodynamic limit. The eigenvector localization exponentially reinforces the infection of highly connected nodes, while it inversely suppresses the infection of small-degree nodes.
\end{abstract}


\maketitle
\section{Introduction}

Networks \cite{network, cn, netS}  essentially capture the structure of individual interactions, contact and mobility patterns,
through which information, innovation, fads, epidemics, and human behaviors spread across us \cite{mobp, behavior, metap, hidden, universal, voter}.
Burst of investigations  have been devoted to SIS model  on networks to understand, predict and control diverse recurrent contagion phenomena, such as influenza-like diseases, computer virus, memes etc. \cite{model,process, unifi}.
There is a threshold $\lambda_{c}$  for the spreading  rate
corresponding to the nonequilibrium absorbing state phase transition \cite{npt, npt1}.
The contagion will outbreak above $\lambda_{c}$,
and  die out below it.
The heterogeneity of real-world networks
characterized by a power law degree distribution $P(k) \sim k^{-\gamma}$
has a nontrivial impact on the threshold.
As a seminal theory to understand SIS model,
the  heterogeneous mean field (HMF) theory
predicts that  for uncorrelated random networks   $\lambda_{c}^{HMF}=\langle k \rangle/\langle k^{2}\rangle$,
where $\langle k \rangle$ and $\langle k^{2} \rangle$ are the first and second moments of $P(k)$ \cite{hmfP}.
For $2<\gamma<3$,  $\langle k^{2} \rangle$ is divergent  in the thermodynamic limit, thus, $\lambda_{c}$ will converge to zero.
While for $\gamma>3$, $\lambda_{c}>0$, since $\langle k^{2} \rangle$ is finite.

In spite of the rather simplicity of SIS model,
considerable controversies and confusions arise from a number of competing theories.
The focus is that $\lambda_{c}>0$ for $\gamma>3$ seems to be a well established fact,
however, different voices  never stop.
The following is a rapid review.
The quenched mean field (QMF) theory replaces the annealed adjacency matrix in HMF by the quenched one, predicting  $\lambda_{c}^{QMF} =1/\Lambda_{1}$,
 where $\Lambda_{1} \approx \max  \{\sqrt{k_{max}}, \langle k^{2}\rangle /\langle k \rangle \} $ is the largest eigenvalue of the adjacency matrix \cite{eigenv, eigenv1}.
 QMF readily gives the same threshold as HMF for $2<\gamma<2.5$.
While for $\gamma>2.5$,
$\Lambda_{1}$ shifts from $\langle k^{2}\rangle /\langle k \rangle$ to $\sqrt{k_{max}}$.
Then  $\lambda_{c}^{QMF} \simeq 1/ \sqrt{k_{max}} $,
which is  zero for any networks with divergent maximum degree \cite{qmf0},
  just at odds with HMF theory.
The node with maximum degree which independently determines $\Lambda_1$ was deemed the self-sustainable source.
Goltsev \emph{et al.}  applied the spectral approach to QMF,
only to find that it causes the  eigenvector localization (EL) \cite{local},
for which  most of the weight of eigenvector  concentrates around
a small number of nodes with highest degree  \cite{el1, el2, el3}.
It immediately  means that there are only a finite number of active nodes  localized around hubs.
Therefore,  $\lambda_{c}^{QMF} \simeq 1/ \sqrt{k_{max}} $ is not a genuine threshold.
The true threshold  appears at the inverse of upper delocalized eigenvalue,
which is slightly smaller than $\lambda_{c}^{HMF}=\langle k \rangle/\langle k^{2}\rangle$.
Once again, it is a positive value for $\gamma>3$.

Lee \emph{et al.}  sparked a big stir \cite{lee}.
It was showed that when $\lambda_{c}^{QMF}< \lambda < \lambda_{c}$,
the activity localized around hubs  in Ref. \cite{local}  in
 fact belongs to the  Griffiths phase \cite{grif}, featuring slow relaxation dynamics.
Owing to the irreversible dynamical fluctuation,
which is seldom considered in most mean-field methods,
these  active domains  eventually  fall into absorbing state
with exceedingly long  relaxation time.
Moreover, they   conjectured that the Griffiths to active phase
 transition  is triggered by the percolation  of the  active
 domains through direct connection of hubs,
which   enables the  mutual reinfection of hubs.
Depending on the corresponding percolation threshold,
the epidemic threshold $\lambda_c$ is also positive for $\gamma>3$.
However, this compelling mechanism is criticized by  Bogu\~{n}\'{a} \emph{et al.}\cite{nature}.
They invoked Durrett's \emph{et al.} work  \cite{durret1, durret2},
 pointing out that the direct connection of hubs is not a necessary condition for reinfection.
The null threshold is retrieved when considering the possibility of reinfection  between any pair of nodes.

Here we argue that these studies involve the key ingredients toward resolving the controversies,
including EL,  dynamical fluctuation and Griffiths effect, the reinfection between distant nodes.
In particular, EL interpolates the Griffiths phase between absorbing  and active phase \cite{gp},
which  may dramatically change our understanding of SIS model.
The  dynamical evolution in this situation cannot be  described
 by those conventional theories  neglecting these points  such as HMF or QMF.
There are  other smart studies without considering  above
ingredients but pairwise dynamical correlations,
such as the dynamical message passing approach \cite{dmp},
effective degree \cite{edegree} and effective branching factor methods \cite{ebf}.
These studies give higher threshold than $\lambda_{c}^{HMF}$,
which are not in agreement with simulations on networks with EL \cite{sample, qs, nature}.

Ref. \cite{nature} seems to put an end to the ongoing debate on the vanishing threshold.
Due to the rather rough annealed mean-field approximation, however,
it provides few  results other than the upper bound  of threshold,
and their  theory   lacks thorough and convincing examination.
Moreover, the  role of EL and Griffiths effect have not been explicitly incorporated in their theory.
And it  remains to be uncovered how the system avoids the irreversible dynamical fluctuation,
triggering  Griffiths to active phase transition.
A consistent and comprehensive investigation of above theoretical ingredients is absent.
We develop a new  mean-field  equation based on Ref. \cite{nature} for this attempt.
It gives us the leverage to examine the theory presented in Ref. \cite{nature}  in several aspects.
It also allows us to study the  impact of EL in depth beyond the framework of QMF.
Our results suggest that SIS model on networks with EL is markedly different from the delocalized case.

\section{Degree-based mean-field approximation }

Let us begin by recalling the theory of Ref. \cite{nature}.
The core is to take into account  that an infected node $i$ may  infect
any other node  $j$ in the network  separated by distance $d_{ij}$  at a long enough time scale at the rate $\bar{\lambda}(d_{ij},\lambda)$.
 ``On long time scales  node $i$ is considered
as susceptible only when the node and all of its nearest
neighbors in the original graph are susceptible'' \cite{nature}.
Therefore, the corresponding recovery rate is replaced by the inverse of the characteristic lifespan of the star $\bar{\delta}(k_{i},\lambda)=\tau^{-1}(k_{i},\lambda)\approx e^{-a(\lambda)k_{i}}$.
The evolution of SIS model is depicted by
\begin{equation}\label{cgmf}
    \frac{d\rho_{i}(t)}{dt}=-\bar{\delta}(k_{i},\lambda)\rho_{i}(t)+(1-\rho_{i}(t))\sum_{j\neq i}\bar{\lambda}(d_{ij},\lambda)\rho_{j}(t),
\end{equation}
where $\bar{\lambda}(d_{ij}\lambda) \approx \lambda \mu^{(d_{ij}-1)}$,
and $\mu= \lambda/(1+\lambda)$ is the probability that an infected node infects its nearest neighbor before returning to susceptible state.
Compared with QMF working on the quenched adjacency matrix,
Eq. (\ref{cgmf}) is defined on a fully connected graph,
where $k_{i}$ and $d_{ij}$ reflect the structure of the original network.
See Appendix \ref{derivation-details} for detailed introduction about Eq. (\ref{cgmf}).

In the following, we develop the degree-based mean-field equation based on Eq. (\ref{cgmf}).
HMF equation is just a result of the degree-based mean-field approximation of QMF equation.
Nonetheless, similar mean-field calculation regarding Eq. (\ref{cgmf}) seems sophisticated.
Incorporating the definition $\rho_{k}(t)\equiv \sum_{deg(i)=k}\rho_{i}(t)/(NP(k))$ with Eq. (\ref{cgmf}),
we obtain
\begin{equation}\label{mf1}
    \frac{d\rho_{k}(t)}{dt}=-\bar{\delta}(k,\lambda) \rho_{k}(t)+(1-\rho_{k}(t))\sum_{k'}\lambda_{kk'} \rho_{k'}(t)NP(k'),
\end{equation}
where $ \lambda_{kk'} \equiv \sum_{deg(i)=k, deg(j)=k'}\bar{\lambda}(d_{ij},\lambda)/NP(k)NP(k')$,
see Eq. (\ref{fkmf})-Eq. (\ref{lkk0}) in Appendix \ref{derivation-details} for the derivations.

A very important mean-field approximation was made in Ref. \cite{nature},
which merely considers the dynamical influence of nodes at the average distance
while omits  others:
\begin{equation}\label{lkk}
    \lambda_{kk'}\approx \bar{\lambda}(d_{kk'},\lambda)=\lambda A_{kk'}^{\ln(1+1/\lambda)/\ln \kappa},
\end{equation}
where $ d_{kk'}=1+\ln(N\langle k\rangle /kk') / \ln \kappa$ is the average distance
 between nodes of degree $k$ and $k'$ for  random networks,
and $\kappa=\langle k^{2} \rangle/\langle k \rangle-1$ is the average branching factor,
 $A_{kk'}=kk'/(N\langle k \rangle)$ is the annealed adjacency matrix for  random networks.

The upper bound of threshold was estimated by plugging Eq. (\ref{lkk}) into  Eq. (\ref{mf1}) in Ref. \cite{nature}.
Though, the rather rough mean-field approximation made in Eq. (\ref{lkk})  impedes further convincing examination of Eq. (\ref{cgmf}).
In particular, the reinfection of a node based on a fully connected  graph
 characterized by the sum $\sum_{j\neq i}\bar{\lambda}(d_{ij},\lambda)\rho_{j}(t)$ lacks faithful validation.
Here we  circumvent above problem by making  a complete consideration of the reinfections of nodes at each distance
\begin{equation}\label{lkk1}
\begin{split}
  \lambda_{kk'}&=\frac{\sum_{\ell=1}^{\infty}\sum_{deg(i)=k, deg(j)=k',  d_{ij}=\ell}\bar{\lambda}(d_{ij},\lambda)}{NP(k)NP(k')}\\
 &=\frac{\sum_{\ell=1}^{\infty}k\kappa^{\ell-1}P(k'|k,\ell)\bar{\lambda}(\ell,\lambda)NP(k)}{NP(k)NP(k')},
  \end{split}
\end{equation}
where $ P(k'|k,\ell) $ is the probability of finding a neighboring
node with degree $ k' $ at distance $ \ell $ from  a given node of degree $ k $.
For sparse uncorrelated random  networks,  $P(k'|k,\ell)=k'P(k')/ \langle k'\rangle$.
Insert it into Eq. (\ref{lkk1}), we have
\begin{equation}\label{lkk2}
  \lambda_{kk'}=\frac{\sum_{\ell=1}^{\infty}\lambda (\mu\kappa)^{\ell-1}kk'}{N\langle k\rangle}= \lambda A_{kk'}\sum_{\ell=1}^{\infty}(\mu\kappa)^{\ell-1}=\Pi\lambda A_{kk'},
\end{equation}
where $\Pi=\sum_{\ell=1}^{\infty}(\mu\kappa)^{\ell-1}$ is a parameter implying the  reinfection patterns.
Numerical simulations show that $\lambda_{c}<\lambda_{c}^{HMF}$ for $\gamma>2.5$ (except for networks with very small size) \cite{nature, qs}, hence,  $\mu\kappa<1$.
Then we have $\Pi=1/(1-\mu\kappa)$.
And we see the ratio of Eq. (\ref{lkk2}) to Eq. (\ref{lkk}),
 $\Pi A_{kk'}^{1-\ln(1+1/\lambda)/\ln \kappa}\gg1$.

Instead of Eq. (\ref{lkk}), substitute Eq. (\ref{lkk2}) into Eq. (\ref{mf1}),
we  obtain
\begin{equation}\label{wei}
     \frac{d\rho_{k}(t)}{dt}=-\bar{\delta}(k,\lambda) \rho_{k}(t)+(1-\rho_{k}(t))\Pi\lambda k\Theta,
\end{equation}
where $\Theta=\sum_{k'}\rho_{k'}(t)k'P(k')/\langle k'\rangle $
 is the probability of a randomly selected link connected to an infected node.
It demonstrates a markedly different physical picture compared with HMF or QMF,
which can be validated by quasi-stationary  simulation \cite{limit, sample} (see Appendix \ref{Quasistationary-simulation} for technical details).

It is worth noting that each term in $\Pi\lambda k\Theta=\sum_{\ell=1}^{\infty}\lambda k\Theta(\mu\kappa)^{\ell-1}$  denoted as $\psi_{k}(\ell) = \lambda k \Theta (\mu \kappa)^{\ell-1}$ describes the reinfections between a node of degree $ k $ and those nodes separated by distance $\ell$.
Let us make it clear.
Obviously, the sum in Eq. (\ref{cgmf}) $\sum_{j\neq i}\bar{\lambda}(d_{ij},\lambda)\rho_{j}(t)=
\sum_{\ell}\sum_{j}\bar{\lambda}(\ell,\lambda)b_{ij}(\ell)\rho_{j}$,
where  $b_{ij}(\ell)=\delta_{d_{ij},\ell}$, and $\delta_{d_{ij},\ell}$ is Kronecker  function.
The degree-based mean-field approximation assumes that all the nodes with degree $k$ are statistically equivalent \cite{process}.
The state of each node of degree $k$ is specified by $\rho_{k}(t)$,
and the quenched network structure is replaced by the average of network ensembles.
Thus, the degree-based mean-field approximation for $\sum_{j\neq i}\bar{\lambda}(d_{ij},\lambda)\rho_{j}(t)$
is $\sum_{\ell} \sum_{k'}\bar{\lambda}(\ell,\lambda)B_{kk'}(\ell)\rho_{k'}(t)NP(k')$.
  Here $B_{kk'}(\ell)=P(k'|k,\ell)k\kappa^{\ell-1}/(NP(k'))$ is the annealed counterpart of $b_{ij}(\ell)$.
As a result, $\psi_{k}(\ell) = \bar{\lambda}(\ell,\lambda) k\kappa^{\ell-1}\sum_{k'} P(k'|k,\ell)\rho_{k'}(t)=\lambda k \Theta (\mu \kappa)^{\ell-1}$,
which decays exponentially with distance.
Detailed derivations for above results are showed by Eq. (\ref{cgmf1})-Eq. (\ref{pipi}) in Appendix \ref{derivation-details}.
As the impact of reinfection at each distance is explicitly taken into account,
Eq. (\ref{wei}) enables us to faithfully verify Eq. (\ref{cgmf}) from a broad perspective.

\section{The exotic critical behavior of $\rho_k$ under the influence of eigenvector localization }

\begin{figure}
  \includegraphics[width=\columnwidth]{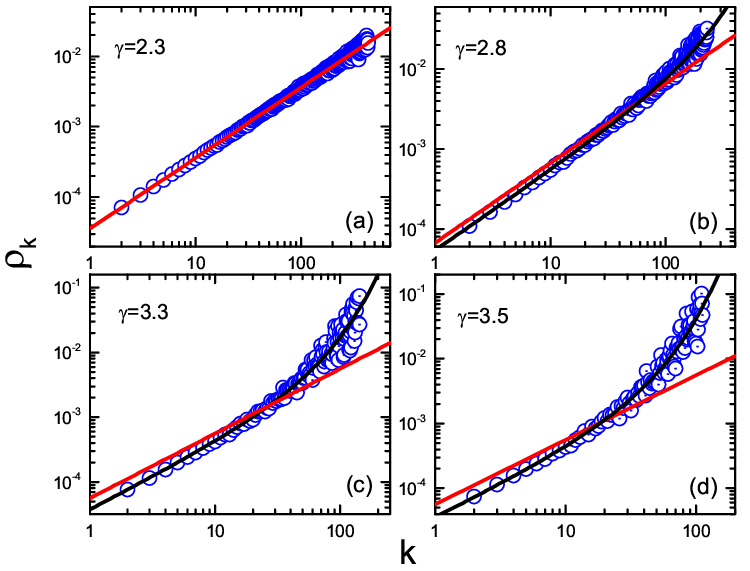}\\
  \caption{(color online) Log-log plot of $\rho_{k}(k)$  at the critical point.
   The  black curve is the prediction of Eq. (\ref{wzw}) (see Appendix \ref{pia} for how the black curve is added),
  while red line is the benchmark $\rho_{k}= \lambda_{c}^{HMF}k\Theta$. In this plot, $N=500000$ and $\lambda_c=0.0272$ for $\gamma=2.3$, $N=1000000$ and $\lambda_c=0.0887$ for $\gamma=2.8$, $N=2000000$ and $\lambda_c=0.1752$ for $\gamma=3.3$, $N=2000000$ and $\lambda_c=0.2175$ for  $\gamma=3.5$.   }\label{Frk}
\end{figure}

First, we study the critical behavior of $\rho_{k}$ around the critical point.
It is not difficult to derive from the HMF  that $\rho_{k}\propto k$.
It seems to be a widely accepted claim.
For example, it was used to estimate the average infection density $\rho$ \cite{lee}.
Indeed, it is true for $2<\gamma<2.5$.
A justified example  is shown in Fig. \ref{Frk} (a).
For $\gamma >2.5$, where EL arises,
however,  it is striking that it no longer holds.
As shown in Fig. \ref{Frk} (b)-(d),  with increasing of $k$,
the deviations from the benchmark get more and more remarkable.
Behind the significant discrepancy lies the manifest failure of the HMF equation.
It calls for new theory.
If Eq. (\ref{lkk}) is plugged into  Eq. (\ref{mf1}),
 we obtain $\rho_k\propto k^{\ln(1+1/\lambda)/\ln \kappa} e^{a(\lambda)k}$ (see Eq. (\ref{bv})-Eq. (\ref{beta}) in Appendix \ref{derivation-details}).
The derivative of this prediction  under double log plot $s(k)\simeq \ln(1+1/\lambda)/\ln \kappa +a(\lambda)k \gg 1$ can explain the huge deviation for the highly connected nodes,
yet fails to predict that $ s(k)\simeq 1 $ for the  small-degree nodes  shown in Fig. \ref{Frk}.

In contrast, the linear approximation of Eq. (\ref{wei}) gives
\begin{equation}\label{wzw}
    \rho_{k}\simeq  \Pi \lambda e^{a(\lambda)k}k\Theta,
\end{equation}
which is perfectly in agreement with simulation, see Fig. \ref{Frk} (b)-(d).
The derivative of  $\rho_{k}$,
\begin{equation} \label{sk}
s(k)= 1 +a(\lambda)k,
\end{equation}
is a coherent explanation for the deviation from benchmark in Fig. (\ref{Frk}).
Assume that $ a(\lambda)\approx a\lambda^{2} $ and $a\sim\mathcal{O}(1)$ \cite{nature},
the deviation $\Delta _{s}=a(\lambda)k$  is pretty tiny for  small-degree nodes,
as it only slightly deviates from $ s(k)=1 $.
Though, with increasing of $ k $,  $\Delta _{s}$ becomes larger and larger.
For highly connected nodes, e.g., $\lambda^{2}k \sim 1$,
the deviation is so huge that by no means can be neglected.
We argue that  the exponential term $e^{a(\lambda)k} $ in Eq. (\ref{wzw}) responsible for the increasing deviation originates from EL.

For random networks, $\Lambda_{1}$ is largely determined by  two special subgraphs in isolation:
the ultramost dense $K$-core \cite{core}  and the  star subgraph composed of the hub with degree $k_{max}$ and its immediate neighbors.
 In specific, $\Lambda_{1}\approx \max \{\Lambda_{1}^{core}, \Lambda_{1}^{star}\}$,
where   $\Lambda_{1}^{core}$ and $\Lambda_{1}^{star}$ are the largest eigenvalues for    $K$-core and the star respectively \cite{eigenv1}.
This insight   suggests that the dynamics are determined by either $K$-core or the star.
When $2<\gamma<2.5$, $\Lambda_{1}\approx \Lambda_{1}^{core}$.
In this case,  the component of the leading  eigenvector is proportional to the degree of the corresponding node.
The dense core, which has a large proportion of the weight of eigenvector \cite{el2, el3},
 plays the role of sustainable  spreading source.
And the spreading can be seen as a branching process.

When $\gamma >2.5$, it turns out that $\Lambda_{1}\approx\Lambda_{1}^{star}$.
More generally, the $n$th largest eigenvalue of the quenched adjacent matrix $\Lambda_{n}$ is almost solely determined by the star centered in  the node with $n$th largest degree $k_{n}$.
For sufficiently large $k_{n}$, $\Lambda_{n}\approx \sqrt{k_{n}}$ \cite{eigenv}.
And the corresponding eigenvector is localized around the center of the star \cite{nlocal}.
Therefore, each star centered in the hub with degree,
e.g., $k \gtrsim 1/\lambda^{2}$, is approximately
dynamically quasi-independent with its own activation
 threshold $\lambda_n \approx 1/\sqrt{k_{n}}$ and lifespan
  $\tau(k,\lambda) \approx e^{a(\lambda)k}$ just like an isolated graph.
Hence, the assumption of $\bar{\delta}(k_i,\lambda)$ is conditional on  EL.
As shown in Fig. \ref{Frk},
Eq. (\ref{wzw}) does not hold for networks with delocalized eigenvector when $\gamma<2.5$.
As a consequence, Eq. (\ref{cgmf}) is invalid when $\gamma<2.5$.

Eq. (\ref{wzw})  can be detailed as
\begin{equation}\label{cdm}
     \sum_{k'}ke^{a(\lambda)k}\frac{k'P(k')}{\langle k\rangle}\rho_{k^{'}}=\frac{1}{\Pi\lambda}\rho_k.
\end{equation}
Based on this result, we define the spreading matrix $\textbf{S}$ with element $S_{kk'}=ke^{a(\lambda)k}k'P(k')/\langle k\rangle$.
Thereby Eq. (\ref{cdm}) is  equivalent to $\textbf{S}\hat{\rho}=\Lambda\hat{\rho}$.
According to the Perron-Frobenius theorem,
if $\Lambda$ is equal to the spectral radius,
then  $\hat{\rho}$ is a positive eigenvector, and vice versa.
Hence, $\Lambda=1/\Pi\lambda$ is a function of the genuine threshold,
marking the onset of endemic state,
where a finite fraction of nodes are infected.
Substitute Eq. (\ref{wzw}) into Eq. (\ref{cdm}), we obtain
a definite critical condition for Griffiths to active phase transition.
\begin{equation}\label{cd}
     \sum_{k=k_{min}}^{k_{max}}\frac{kP(k)}{\langle k\rangle}\Pi\lambda k e^{a(\lambda)k}=1,
\end{equation}
where $k_{min}$ denotes the minimum degree.

It is interesting  that Eq. (\ref{cd}) means that $\textmd{Tr}(\textbf{S})=\Lambda_{1}$,
where  $\textmd{Tr}(\cdot)$ is the trace operator.
Similar result also holds for  the annealed adjacency matrix $A_{kk'}$,
that is, $\textmd{Tr}(\textbf{A})=\langle k^{2}\rangle/\langle k\rangle$.
It is not surprising. Owing to $[\textmd{Tr}(\textbf{S})]^{2}=\textmd{Tr}(\textbf{S}^{2})$
and $[\textmd{Tr}(\textbf{A})]^{2}=\textmd{Tr}(\textbf{A}^{2})$,
there is only one non-zero eigenvalue for the matrix.

Since $e^{a(\lambda)k_{min}}\sum_{k} k^{2}P(k) <\sum_{k} k^{2}P(k) e^{a(\lambda)k} <e^{a(\lambda)k_{max}}\sum_{k} k^{2}P(k)$, there must exist $k_x\in(k_{min},k_{max})$  making
$\sum_{k} k^{2}P(k) e^{a(\lambda)k}=\langle k^{2}\rangle e^{a(\lambda)k_{x}}$.
With this result,  Eq. (\ref{cd}) is equivalent to,
\begin{equation}\label{th}
  \Pi_{c} \lambda_{c} e^{a(\lambda_{c})k_{x}}=\lambda_{c}^{HMF},
\end{equation}
where $\Pi_{c}$ is the corresponding value of $\Pi$ at the critical point.
Combine Eq. (\ref{th}) with Eq. (\ref{wzw}),
in the vicinity of critical point,
 the later can be written as
\begin{equation}\label{rs}
\rho_k\simeq \lambda_{c}^{HMF}k\Theta e^{(k-k_x)a(\lambda_{c})},
\end{equation}
which has appealing implications.
EL exponentially reinforces the infection of large-degree nodes ($k>k_x$), while  inversely suppresses the infection of small-degree nodes ($k<k_x$), see Fig. \ref{Frk}.

\section{Examining the value of $\Pi$ and estimation of the threshold for Griffiths to active phase transition}
\begin{figure}
  \includegraphics[width=\columnwidth]{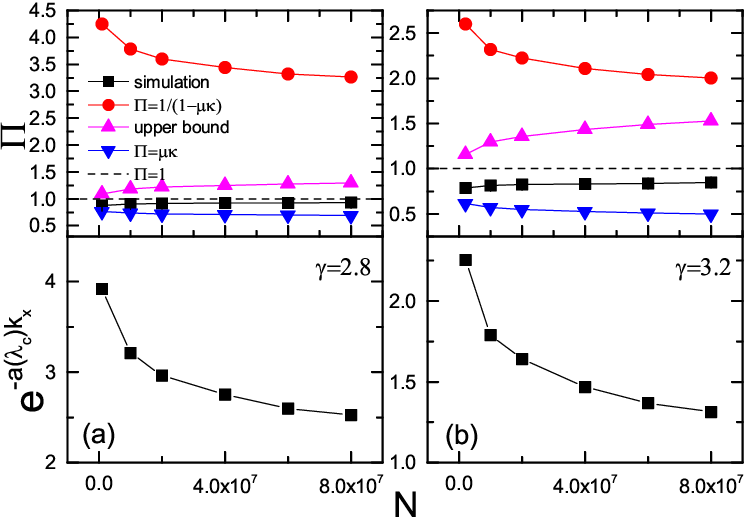}\\
 \caption{(color online)  Examination of the value of $\Pi$.
  Up panel:  value of $\Pi$ calculated based on simulated data compared with  $\Pi=1/(1-\mu\kappa)$, $\Pi=\mu\kappa$, and the upper bound given by Eq. (\ref{upb}).
  Bottom panel: computation of $e^{-a(\lambda_{c})k_{x}}$ via plugging $\Pi=1/(1-\mu\kappa)$ and the simulated value of $\lambda_{c}$ into Eq. (\ref{th}).
}\label{PI}
\end{figure}

There is a big drawback for the rough mean-field approximation made in Eq. (\ref{lkk}).
 It is unable to confirm the validity of  the sum  in  Eq. (\ref{cgmf})
 characterizing the reinfection patterns defined on a fully connected graph.
Eq. (\ref{wei}) suggests that $\Pi$  can serve as an effective indicator.
It can be best verified by checking  whether $\Pi=\sum_{\ell=1}^{\infty}(\mu\kappa)^{\ell-1}=1/(1-\mu\kappa)$.
We provide  three approaches to  examine  this theoretical prediction.

On one hand, there are two parameters $\Pi$ and $a$,
the values of which remain to be determined.
On the other hand, we have two equations (Eq. (\ref{wzw}) and Eq. (\ref{cd})) involving the two parameters.
This mathematically allows us to solve the two parameters with assistance of the simulation data of $\lambda_{c}$  and $\rho_k$,
see Appendix \ref{pia} for how this is done.
As shown in Fig. \ref{PI}, the corresponding numerical solutions  suggest that $\Pi<1$,
which is significantly smaller than  $\Pi=\sum_{\ell=1}^{\infty}(\mu\kappa)^{\ell-1}>1$.

Besides, Eq. (\ref{th})  provides another two methods to verify the theoretical value of $\Pi$.
Plug $\Pi=1/(1-\mu\kappa)$ and the simulated value of $\lambda_{c}$ into Eq. (\ref{th}),
$e^{-a(\lambda_{c})k_{x}}=\lambda_{c}\Pi/\lambda_{c}^{HMF}$.
It unexpectedly produces an invalid result $e^{-a(\lambda_{c})k_{x}}>1$,
see Fig. \ref{PI} for demonstration.
In contrast, $e^{-a(\lambda_{c})k_{x}}<1$ always hold if $\Pi<1$ and $\lambda_{c}<\lambda_{c}^{HMF}$ .

Utilizing Eq. (\ref{th}) and the strong constraint $\lambda_{c}<\lambda_{c}^{HMF}$,
we obtain the upper bound  of $\Pi$.
Since $e^{-a(\lambda)k_{x}}<1$,
we  have
\begin{equation}\label{upb}
  \Pi < \frac{\lambda_{c}^{HMF}}{\lambda_{c}}.
\end{equation}
Since $\lambda_{c}$ can be faithfully obtained from simulation,
this upper bound is useful for finite size systems.
As displayed in Fig. \ref{PI},  the theoretical prediction  $\Pi=1/(1-\mu\kappa)$ is substantially bigger than this upper bound, which thus violates the inequality of Eq. (\ref{upb}).
However, the simulation value $\Pi_{s} <1$  well obeys this inequality .

Now we are prepared to analyze the threshold $\lambda_c$.
Similar to Ref. \cite{nature},
the upper bound of  $\lambda_c$ can be estimated by merely
 considering the reinfections from those nodes at the average distance,
since $\Pi\lambda \gg \bar{\lambda}(\langle\ell\rangle,\lambda) $,
where $\langle\ell\rangle \sim \ln N$ is the average distance between any two nodes in a small-world network.
The upper bound of $\lambda_{c}$ is obtained by substituting $\bar{\lambda}(\langle\ell\rangle,\lambda)$ for $\Pi\lambda$ in Eq. (\ref{cd}).
Due to $\sum_{k} k^{2}P(k)e^{a(\lambda)k}/\langle k\rangle>\sum_{k} kP(k)e^{a(\lambda)k}/\langle k\rangle=\langle \tau\rangle$,
the upper bound can be estimated by
$\bar{\lambda}(\langle\ell\rangle,\lambda)\langle \tau\rangle=1$,
or $1/\bar{\lambda}(\langle\ell\rangle,\lambda)=\langle \tau\rangle$.
Here, $1/\bar{\lambda}(\langle\ell\rangle,\lambda)$ is just the average time it takes
for an infected node to infect for the first time a node at the average distance 
Hence, it also provides a condition that on average a star with an active center is capable of propagating
the infection to nodes as far as at the average distance before spontaneously recovering.
It requires $\langle \tau\rangle\rightarrow\infty$ in the thermodynamic limit.
Without compromising the upper bound,
$\bar{\lambda}(\langle\ell\rangle,\lambda)\langle \tau\rangle=1$ can be reduced to
 \begin{equation}\label{upperbound}
   \langle k\rangle=e^{a(\lambda)k_{max}+\ln k_{max}+\ln P(k_{max})+ \ln N \ln\lambda}.
 \end{equation}

If $k_{max}$ grows faster than  $\ln N$,
and $P(k)$ decays  slower than exponential,
 with increasing of $N$ while keeping $\lambda$ fixed,
then the exponential term will become bigger than $\langle k\rangle$.
As a result, $\lambda$ must converge to zero in the thermodynamic limit.
Assume that $a(\lambda)\propto \lambda^{2}$,
we conclude that  $\lambda_{c} \rightarrow 0$ at a speed
much slower than $\lambda_{c}^{QMF} \simeq 1/ \sqrt{k_{max}} $ for $\gamma >2.5$.

\begin{figure}
  \includegraphics[width=\columnwidth]{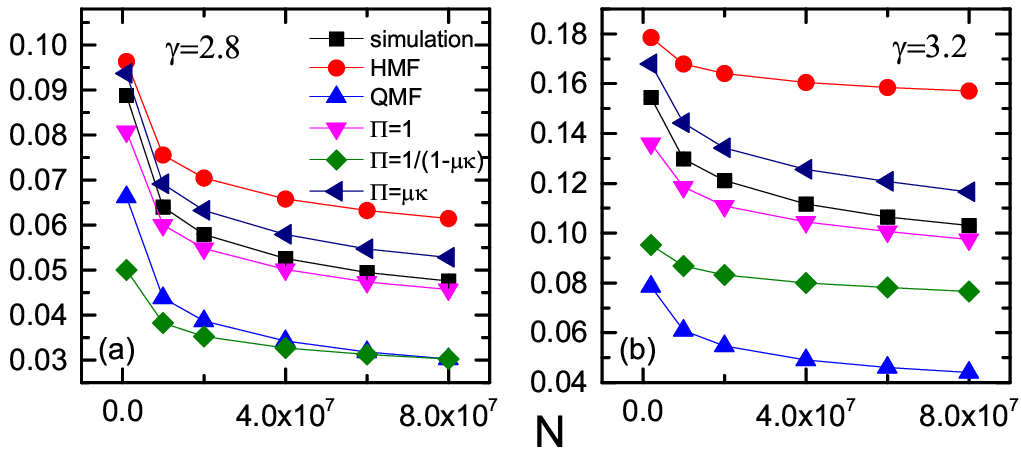}\\
 \caption{(color online) Comparison of the threshold (showed by the vertical axis) calculated based on Eq. (\ref{cd}) with that of simulation and HMF, QMF.
}\label{lc}
\end{figure}

Quasi-stationary simulations have already shown that
$\lambda_{c}^{QMF}<\lambda_c<\lambda_{c}^{HMF}$  for $\gamma>2.5$ \cite{sample, qs, nature}.
Here we  provide closer bounds for $\lambda_c$ by substituting the corresponding
bound of $\Pi$ into Eq. (\ref{cd}).
The vanishing threshold means $\mu\kappa \rightarrow 0$ for $\gamma>3$,
while the simulated value  $\Pi_{s}\rightarrow 1$.
As a result, for large-size networks, $\mu\kappa<\Pi_{s}<1$ for $\gamma>3$.
For $2.5<\gamma<3$, Fig. \ref{PI} shows that it is true as well.
Therefore, a closer upper bound for $\lambda_c$ can be obtained if  $\Pi=\mu\kappa$,
and $\Pi=1$  accordingly leads to a closer lower bound.
In contrast, inserting $\Pi=1/(1-\mu\kappa)$ into
Eq. (\ref{cd})  yields a  threshold much smaller than simulation,  see Fig. \ref{lc}.

\section{Discussion and conclusion}
Our results allow us to evaluate to what extent Ref. \cite{nature} is valid.
First, the introduction of $\delta(k_i,\lambda)$ is closely associated with  EL.
It is invalid when $\gamma<2.5$.
However, Ref. \cite{nature}  claimed that Eq. (\ref{cgmf}) can be applied to any network.
The second concern is the remarkable discrepancy between
 the  value of $\Pi$ obtained from simulation and the  mean-field expectation $\Pi=1/(1-\mu\kappa)$,
which stems from  the sum in  Eq. (\ref{cgmf}) defined on a fully connected graph
portraying  the pattern of mutual reinfections between distant nodes
accounting for the null threshold.
This remarkable contradiction underscores that Eq. (\ref{cgmf}) is inadequate
to characterize the dynamical process.
Since it is difficult to reconcile this confusing contradiction,
We leave it as an open question.

Although Eq. (\ref{wei}) does not have complete theoretical power as the correct  expression for $\Pi$ remains unclear, it  has already produced a series of novel results,
answering some important questions of broad interest.
Eq. (\ref{cd}) portrays the critical condition for Griffiths to active phase transition: on average,  a node getting infected  can  further  infect another node within the lifespan of the star centered in itself.
Thereby, the star  is able to avoid getting trapped in the absorbing state,
marking the onset of the phase transition.
As a hallmark of the transition,
we show how the critical behavior of $\rho_k$ is dramatically changed by   EL.
Instead of the conventional belief $\rho_k\propto k$,
it is striking that $\rho_k \propto k e^{a(\lambda)k}$ for $\gamma >2.5$, where the exponential term originates from EL.
As a consequence,  HMF and QMF   are not valid  when  $\gamma >2.5$.

On one hand, Eq. (\ref{upperbound}) reproduces the conclusion
of Ref. \cite{nature} regarding the vanishing threshold.
On the other hand, it suggests that whether $\lambda_{c} \rightarrow 0$
depends on the competition between  the damping long-range
reinfection rate $\bar{\lambda}(\langle\ell\rangle,\lambda)$
and  the  growing average lifespan $\langle \tau\rangle$.
The divergence of $\langle \tau\rangle$ is a crucial condition
for the effective mutual reinfection between remote nodes
and the null threshold in the thermodynamic limit.
Eq. (\ref{th}) makes a direct connection between $\lambda_{c}$ and $\lambda_{c}^{HMF}$.
It could be seen as a  quantitative criterion  explicitly separating highly connected nodes ($k>k_x$)
from  small-degree nodes ($k<k_x$).
Together with Eq. (\ref{rs}),  we conclude that owing to  EL,
the infection of highly connected nodes is exponentially reinforced,
while  the infection of small-degree nodes is inversely suppressed.





\textbf{Acknowledgement.} We thank S. C. Ferreira for the discussion of quasistationary simulation.
We are also grateful to  Jia-Rong Xie for  discussing the manuscript.
This work is funded by the National Natural Science Foundation
of China (Grant No. 71874172).

\appendix

\section{Some derivation details for the degree-based mean-field equation and threshold}
\label{derivation-details}

If the dynamical correlations are neglected,
the quenched mean field (QMF) equation takes the form:
\begin{equation}\label{qmf}
  \frac{d\rho_{i}(t)}{dt}=-\rho_{i}(t)+\lambda (1-\rho_{i}(t)) \sum_{j}a_{ij}\rho_{j}(t),
\end{equation}
where $\rho_{i}(t)$ is the infection density of node $i$,
and  $a_{ij}$  is the element of adjacency matrix.
Next, we develop the degree-based mean-field equation based on QMF.
The degree-based mean-field approximation basically assumes that all the nodes with the same degree are statistically equivalent.
This assumption has two implications.
All the  nodes of degree $k$  share the same sate $\rho_{k}(t)$.
And the quenched  adjacency matrix $a_{ij}$ is replaced by the average of network ensembles.
\begin{equation}\label{qam}
  A_{k_ik_j}=\frac{P(k_j|k_i)k_i}{NP(k_j)}.
\end{equation}
Applying above two  principles to Eq. (\ref{qmf}),
the resulting degree-based mean-field equation is
\begin{equation}\label{hmf}
   \frac{d\rho_{k}(t)}{dt}=-\rho_{k}(t)+\lambda(1-\rho_{k}(t))\sum_{k'} A_{kk'}\rho_{k'}(t)NP(k').
\end{equation}
Plug $A_{kk'}=kk'/(N\langle k\rangle)$ for uncorrelated random networks
into Eq. (\ref{hmf}), we obtain the familiar form of HMF equation:
\begin{equation}\label{hmf1}
   \frac{d\rho_{k}(t)}{dt}=-\rho_{k}(t)+(1-\rho_{k}(t))\lambda k\Theta,
\end{equation}
where $\Theta=\sum_{k'}\rho_{k'}(t)k'P(k')/\langle k'\rangle $
refers to the probability of a randomly selected link pointing to an infected node.
The linear approximation of  Eq. (\ref{hmf1}) around the critical point leads to
\begin{equation}\label{rk}
\rho_{k}\simeq \lambda k\Theta .
\end{equation}

The QMF equation only involves the influence of the nearest neighbors.
In contrast, Bogu\~{n}\'{a} \emph{et al}. took a leap by
considering that at a long enough time scale, a node suffers  the infections of all the nodes in the network  \cite{nature}.
The spreading rate $\lambda$ is replaced by the effective rate
\begin{equation}\label{lata}
    \bar{\lambda}(d_{ij},\lambda) \approx \lambda e^{-b(\lambda)(d_{ij}-1)}=\lambda \mu^{(d_{ij}-1)},
\end{equation}
where $b(\lambda)=\ln(1+1/\lambda)$, $\mu=\lambda/(1+\lambda)$.
It denotes the rate at which an infected node $i$ can infect a node $j$
separated by distance $d_{ij}$.
``On long time scales  node $i$ is considered
as susceptible only when the node and all of its nearest
neighbors in the original graph are susceptible'' \cite{nature},
the recovery rate of node $i$ is then replaced by the  inverse of the characteristic lifetime of star subgraph composed of the node and its nearest neighbors: $\bar{\delta}(k_{i},\lambda)=\tau^{-1}(k_i,\lambda)\approx e^{-a(\lambda)k_{i}}$,
where $a(\lambda)\propto \lambda^{2}$.
Based on these points,
they suggested a time coarse-grained equation, i.e., Eq. (\ref{cgmf}).
Combining $\rho_{k}(t)\equiv \sum_{deg(i)=k}\rho_{i}(t)/(NP(k))$ with Eq. (\ref{cgmf}),
we obtain
\begin{equation}\label{fkmf}
\begin{split}
  &\frac{d\rho_{k}(t)}{dt}=-\bar{\delta}(k,\lambda) \rho_{k}(t)+ \\
  &(1-\rho_{k}(t)) \frac{\sum_{deg(i)=k} \sum_{j\neq i}\bar{\lambda}(d_{ij},\lambda)\rho_{j}(t)}{NP(k)}.
\end{split}
\end{equation}

The sum in this equation can be further calculated as follows:
\begin{equation}\label{sum}
\begin{split}
  &\frac{\sum_{deg(i)=k} \sum_{j\neq i}\bar{\lambda}(d_{ij},\lambda)\rho_{j}(t)}{NP(k)}\\
  &=\frac{\sum_{k'} \sum_{deg(i)=k,deg(j)=k'}\bar{\lambda}(d_{ij},\lambda)\rho_{j}(t)}{NP(k)}\\
  &=\frac{\sum_{k'} \rho_{k'}(t)\sum_{deg(i)=k,deg(j)=k'}\bar{\lambda}(d_{ij},\lambda)}{NP(k)}\\
  &=\sum_{k'}NP(k') \rho_{k'}(t)\frac{\sum_{deg(i)=k,deg(j)=k'}\bar{\lambda}(d_{ij},\lambda)}{NP(k)NP(k')}\\
  &=\sum_{k'}NP(k') \rho_{k'}(t)\lambda_{kk'},
\end{split}
\end{equation}
where
\begin{equation}\label{lkk0}
 \lambda_{kk'}=\frac{\sum_{deg(i)=k, deg(j)=k'}\bar{\lambda}(d_{ij},\lambda)}{NP(k)NP(k')}
\end{equation}

The mean-field calculation of $\lambda_{kk'}$ becomes crucial.
Ref.\cite{nature} coped  with this problem by making an important mean field approximation
which merely keeps the contribution of nodes at the average distance
while omits  others: $\lambda_{kk'}\approx \bar{\lambda}(d_{kk'},\lambda)$
where $ d_{kk'}=1+\ln(N\langle k\rangle /kk') / \ln \kappa$ is the average distance
 between nodes of degree $k$ and $k'$ for small-world random networks.
It leads to a rather sophisticated equation reported in Ref.\cite{nature}:
\begin{equation}\label{bv}
\begin{split}
  &\frac{d\rho_{k}(t)}{dt}=-\bar{\delta}(k,\lambda) \rho_{k}(t)+ \\
  &\lambda N \left[ \frac{k}{N \langle k
     \rangle}\right]^{\frac{b(\lambda)}{\ln{\kappa}}}
  \sum_{k'} k'^{\frac{b(\lambda)}{\ln{\kappa}}}P(k')
  \rho_{k'}(t)[1-\rho_k(t)].
\end{split}
\end{equation}

In the vicinity of critical point, Eq. (\ref{bv}) yields
a prediction significantly different from  HMF theory (see Eq. (\ref{rk})):
\begin{equation}\label{bv1}
\rho_k\simeq e^{a(\lambda)k}\lambda N \left[ \frac{k}{N \langle k
     \rangle}\right]^{\frac{b(\lambda)}{\ln{\kappa}}}
  \sum_{k'} k'^{\frac{b(\lambda)}{\ln{\kappa}}}P(k')
  \rho_{k'}
\end{equation}
In short,
\begin{equation}\label{beta}
    \rho_k\propto k^{\beta} e^{a(\lambda)k},
\end{equation}
where $ \beta=b(\lambda)/ \ln \kappa $,
and $\kappa=\langle k^{2} \rangle/\langle k \rangle-1$ is the average branching factor.

We propose a different degree-based mean-field approximation of Eq. (\ref{cgmf}),
which can be rewritten as
\begin{equation}\label{cgmf1}
    \frac{d\rho_{i}(t)}{dt}=-\bar{\delta}(k_{i},\lambda)\rho_{i}(t)+
    (1-\rho_{i}(t))\sum_{\ell}\sum_{j}\bar{\lambda}(\ell,\lambda)b_{ij}(\ell)\rho_{j},
\end{equation}
where  $b_{ij}(\ell)=\delta_{d_{ij},\ell}$.
According to the principles of degree-based mean-field approximation,
the sum in above equation is replaced by
\begin{equation}
\sum_{\ell} \sum_{k'}\bar{\lambda}(\ell,\lambda)B_{kk'}(\ell)\rho_{k'}(t)NP(k'),
\end{equation}
where $B_{kk'}(\ell)=P(k'|k,\ell)k\kappa^{\ell-1}/(NP(k'))$ is the annealed counterpart of $b_{ij}(\ell)$,
and $ P(k'|k,\ell) $ is the probability of finding a neighboring
node with degree $ k' $ at distance $ \ell $ from  a given node of degree $ k $.
For sparse uncorrelated random  networks,  $P(k'|k,\ell)=k'P(k')/ \langle k'\rangle$.
As a result,
\begin{equation}\label{path}
  \begin{split}
    &\sum_{\ell} \sum_{k'}\bar{\lambda}(\ell,\lambda)B_{kk'}(\ell)\rho_{k'}(t)NP(k')\\
    &=\sum_{\ell} \bar{\lambda}(\ell,\lambda) k\kappa^{\ell-1}\sum_{k'} P(k'|k,\ell)\rho_{k'}(t)\\
    &=\lambda k \Theta\sum_{\ell} (\mu \kappa)^{\ell-1}\\
    &=\Pi\lambda k\Theta,
  \end{split}
\end{equation}
where
\begin{equation}\label{pipi}
  \Pi=\sum_{\ell} (\mu \kappa)^{\ell-1}=\frac{1}{1-\mu \kappa}.
\end{equation}
Since the influence of nodes at each distance is completely taken into account,
we derive the least approximated degree-based mean-field equation for Eq. (\ref{cgmf}),  i.e.,  Eq. (\ref{wei}).
It enables us to understand  Griffiths to active phase transition.





\section{Quasistationary  simulation}
\label{Quasistationary-simulation}
In this paper, we use the configuration model to generate
uncorrelated random networks for quasistationary simulation \cite{ucm}.
The degree sequence $\{k_i\}$ is generated as follows.
First, the degree distribution $P(k)=ck^{-\gamma}$ is normalized as
$\sum_{k_{min}}^{k_{c}} P(k)dk=1$, where $k_c \leq  \sqrt{N}$
is the structural degree cutoff that can avoid inherent degree correlation \cite{ucm},
 and $k_{min}$ refers to the minimum degree.
Second, the number of nodes with degree $k$ is $n(k)=NP(k)$.

The principle of  quasistationary simulation is to calculate the statistical average
over possible evolutionary configurations by certain sampling strategy without being trapped in the absorbing state.
The discrete-time approach, which synchronously  updates the states of all the nodes with a fixed time interval such as $\Delta t=1$,
can lead to incorrect result,
especially when the recovery rate is unit \cite{limit}.
Instead, the continuous-time approach is employed in this paper.

\begin{figure}
  \includegraphics[width=\columnwidth]{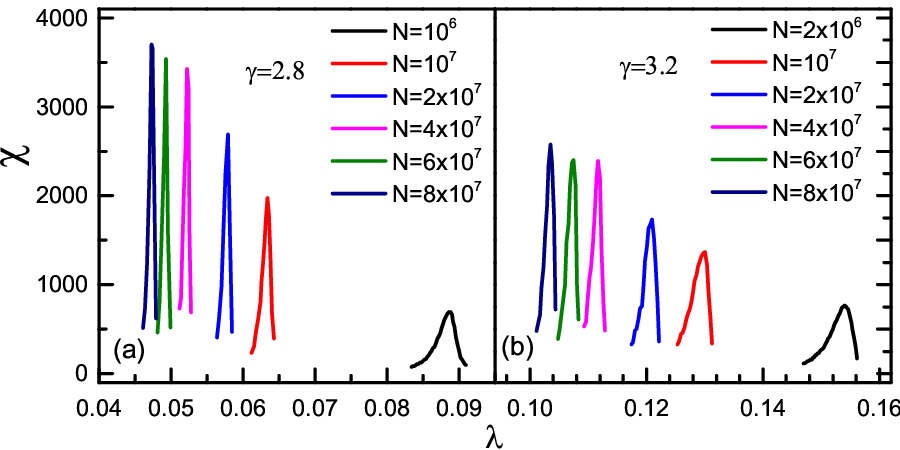}\\
 \caption{(color online) Examples of the susceptibility curve produced by quasi-stationary   simulation.
}\label{x}
\end{figure}

It is implemented according to the following algorithm.
At each time step, the number of infected nodes $N_{i}$ and the sum of the degree of these nodes $E_{i}$ are computed respectively.
With probability $N_i/(N_i+\lambda E_i)$, a randomly picked infected node returns to susceptible state. While with probability $\lambda E_i/(N_i+\lambda E_i)$,
an edge connected to an infected node is chosen at random.
If another endpoint of the edge is a susceptible node,
the node is set as infected. Otherwise, the simulation just continues.
After this operation, recompute the values of $N_{i}$ and $E_{i}$,
and the total time increases by the amount  $\Delta t=1/(N_i+\lambda E_i)$.
Above procedures are iterated.
The threshold is determined by the position of  the peak in the susceptibility curve.
And the susceptibility is defined as
\begin{equation}\label{susc}
  \chi=\frac{\langle n^2\rangle -\langle n\rangle^2}{\langle n\rangle}=N\frac{\langle\rho^2\rangle-\langle\rho\rangle^2}{\langle\rho\rangle}.
\end{equation}

Once the system visits absorbing state,
how to select the  seeds  to restart the evolution is crucial.
It determines whether the threshold can be accurately captured.
The standard updating scheme is  to randomly select a historical evolution configuration from a costly maintained memory.
Such scheme is not a good one,
since it does not produce a susceptibility curve with sharp peak,
but multiple  ambiguous peaks.
In this case, it is difficult to confirm the accurate position of the transition point.
That is why Ref. \cite{error} provided misleading results.
Another scheme which fixes the hub node as restarting  seed also suffers such a drawback \cite{sample}.
Instead,  we employ the so called reflection boundary condition,
which simply selects the last active node as the restarting  seed \cite{sample}.
This method is able to overcome above problem.
See Fig. \ref{x} for examples.

\begin{figure}
  \includegraphics[width=\columnwidth]{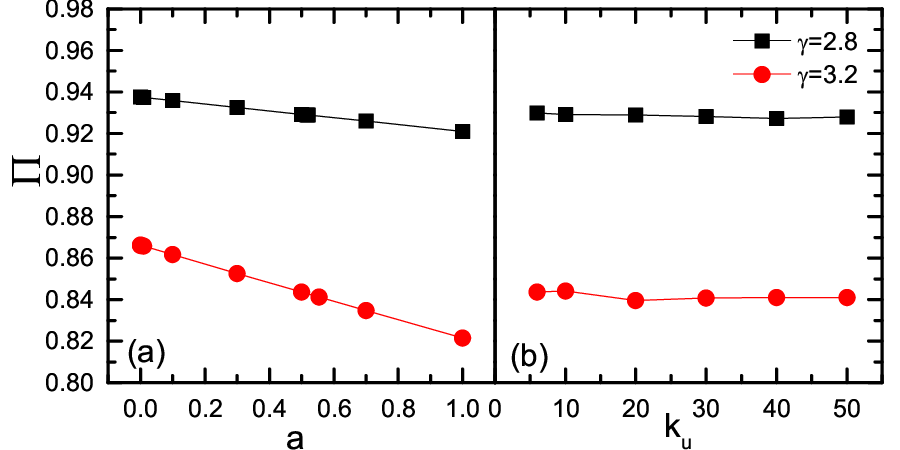}\\
 \caption{(color online) The influence  of the values of $a$ and $k_{u}$ to measure $\Pi$.
    (a) $k_u=10$, (b) $a=0.5$. In this example, $N=80000000$ for both networks. And the threshold $\lambda_c=0.0473$ for $\gamma=2.8$,  $\lambda_c=0.1038$ for $\gamma=3.2$.
}\label{pi}
\end{figure}

\section{How to determine the values of $\Pi$ and $a$}
\label{pia}
The sum over infections of all the other nodes in Eq. (\ref{cgmf})  is defined on a fully connected graph, which  highlights its difference compared  with QMF theory.
The validity of this theoretical proposition remains to be confirmed.
It can be effectively addressed by  examining whether $\Pi=\sum_{\ell=1}^{\infty}(\mu\kappa)^{\ell-1}=1/(1-\mu\kappa)$.
The quasi-empirical parameter $a(\lambda)$
in  Eq. (\ref{wzw}) makes it a nontrivial job.
Though, the combination of Eq. (\ref{wzw}) and Eq. (\ref{cd}) involving the two parameters mathematically allows us to determine their values using the simulation data of $\lambda_{c}$  and $\rho_k$.
By fitting the data of $\rho_k$ with Eq. (\ref{wzw}), we can roughly estimate their values.
Meanwhile, only the values of $\Pi$ and $a$ satisfying Eq. (\ref{cd}) can be accepted.
It is not difficult to find the valid values by iterative computations.
In the following, let us clarify how this works.

Eq. (\ref{wzw}) suggests that the derivative of $\rho_{k}$ under double log plot is $s(k)=1+a(\lambda)k$.
Assume that $a(\lambda)\approx a\lambda^{2}$ \cite{nature},
and $a\sim \mathcal{O}(1)$,
\begin{equation}\label{s}
  s(k)\approx 1+ a\lambda^{2}k,
\end{equation}
which is well in agreement with simulated results.
Note that if $a\gg1$,  $s(k)$ seriously deviates from the benchmark $s(k)=1$ for the small-degree nodes. And if $a\rightarrow 0$, the deviation from the benchmark for highly connected nodes will be so tiny that can be neglected . Both results do not coincide with simulation. Therefore, it is reasonable to assume that $a\sim \mathcal{O}(1)$.
See the main text for detailed illustration.
For  small degree nodes, since $a\lambda^{2}k\ll1$,  Eq. (\ref{wzw}) can be approximated as
\begin{equation}\label{mp}
  \rho_k \approx  \Pi\lambda k\Theta(1+a\lambda^{2}k).
\end{equation}
In this case, the value of $a$ has little influence  to estimate $\Pi$.
Thus, it is  plausible to just try a specific value from  $0<a\leq 1$,
and $\Pi$ is estimated by fitting the data of $\rho_k$ constrained
by $k<k_u$ with Eq. (\ref{wzw}).
 Figure. \ref{pi} shows that both the values of $a$ and $k_u$ do have little influence  to estimate $\Pi$.
With increasing  of $a$, as shown in Fig. \ref{pi} (a),
 $\Pi$ decreases slowly in a linear way.

\begin{figure}

  \includegraphics[width=\columnwidth]{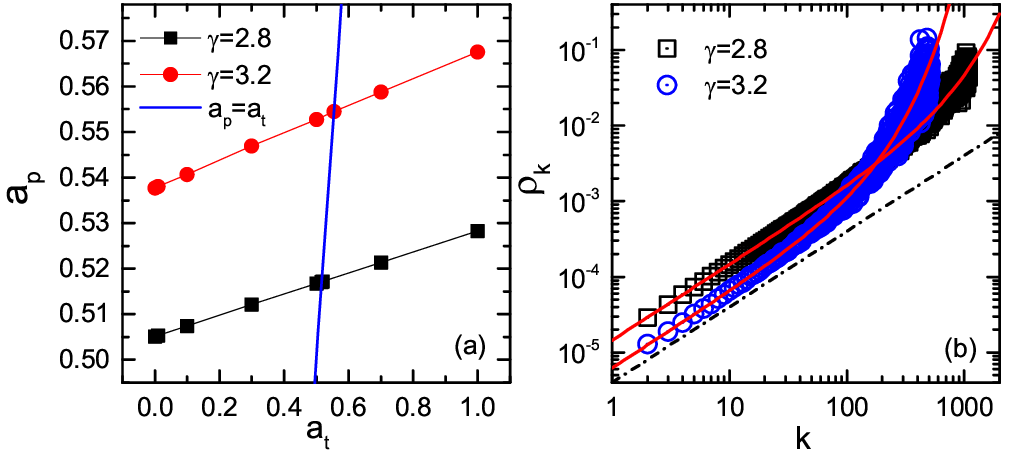}\\
  \caption{(a) The trial value of $a$ denoted as $a_t$ versus the predicted value of $a$ represented by $a_p$. The cross point is assigned to $a$.
    (b) Validation of Eq. (\ref{wzw}). The red curves are prediction of Eq. (\ref{wzw}) with $a$ and $\Pi$ determined by iterative computations.
    The  dash line with slope $s=1$ is  added for benchmark.
    The networks here are the same as those in Fig.  \ref{pi}}\label{a}.
\end{figure}

The values of $\Pi$ and $a$ can be further determined in a consistent way.
Fit the data of $\rho_k$  with  a trial value of $a$ denoted as $a_t$ to  yield the value of $\Pi$.
Eq. (\ref{cd}) puts a strong constraint on $\Pi$ and $a$.
Hence, substitute this value of $\Pi$ into Eq. (\ref{cd}),
 and obtain $a_p$ referring to the predicted value of $a$.
Once $a_p$ is worked out, give this value to $a_t$.
Iterate above computation procedure for several times.
It quickly converges to $a_p=a_t$, see Fig. \ref{a} (a).
And  $\Pi$ is  accordingly determined as well.
Eq. (\ref{wzw}) is then specified by substituting the values of $a$ and $\Pi$.
As shown in Fig. \ref{a} (b), the specified expression of
 Eq. (\ref{wzw}) fits fairly well with the simulated data.
Finally, the fidelity of above computation is enough to come to the conclusion   $\Pi<1$.


\end{document}